\documentclass[doublecol,final]{epl2}

\usepackage[latin1]{inputenc}
\usepackage{amsmath,amssymb,bbm,epsfig}

\makeatletter\renewcommand{\epl@stylemark}{\hbox to0pt{%
\hskip0em\vbox to 0pt{\vss\hbox{\sffamily
 published in EPL \textbf{77}, 50007 (2007)}\vskip6ex}\hss}}
\makeatother

\newcommand{\assign}{:=}

\newcommand{\bignone}{\,}

\newcommand{\mathd}{\mathrm{d}}

\newcommand{\tmmathbf}[1]{\ensuremath{\boldsymbol{#1}}}
\newcommand{\tmop}[1]{\ensuremath{\operatorname{#1}}}

\title{Monitoring approach to open quantum dynamics\\ using scattering
theory}
\shorttitle{Monitoring approach to open quantum dynamics}
\author{Klaus Hornberger}
\institute{Arnold Sommerfeld Center for Theoretical Physics,\\
Ludwig-Maximilians-Universität München,
Theresienstraße 37, 80333 Munich, Germany}
\date{\today}
\pacs{03.65.Yz}{Decoherence; open systems; quantum statistical methods}
\pacs{03.65.Nk}{Scattering theory}
\pacs{34.10.+x}{General theories and models of atomic and molecular collisions and interactions}

\begin{document}
\abstract{
  It is shown how S-matrix theory and the concept of continuous quantum
  measurements can be combined to yield Markovian master equations which
  describe the environmental interaction non-perturbatively. The method is
  then applied to obtain the master equation for the effects of a gas on the
  internal dynamics of an immobile complex quantum system, such as a trapped
  molecule, in terms of the exact multi-channel scattering amplitudes.
}
\maketitle

\section{Introduction}

A truism of quantum physics tells that no system is perfectly isolated and it
is therefore not surprising that the study of open quantum systems is an
ubiquitous \ theme of present day quantum mechanics, see
{\cite{Breuer2002a,Carmichael1993a,Gardiner2000a,Weiss1999a}} and refs.
therein. An important class of evolution equations for open systems are
Markovian master equations. They imply that environmental correlations
disperse fast, so that on a coarse-grained timescale the temporal change of
the system state $\rho$ depends on the present state of the system, but not on
its history. From the strict point of view of an operationalist (who dismisses
the notion of a ``system'' altogether and takes $\rho$ as describing an
equivalence class of preparations in the lab {\cite{Kraus1983a}}) one may even
argue that any valid differential equation for the time evolution of $\rho$
must generate a completely positive dynamical semigroup {\cite{Alicki1987a}}
and must hence be Markovian.

Putting the pros and cons of Markovian vs. non-Markovian formulations aside,
it is fair to say that a large class of open quantum systems is described
appropriately by time-local master equations. At the same time, it is curious
that the Markov property does not emerge naturally in standard microscopic
derivations. Rather, one has to impose it ``by hand'', usually by interpreting
some quantities as correlation functions, which must then be assumed to be
$\delta$-correlated. This may be still transparent in weak coupling
calculations such as the Bloch-Redfield \ approach {\cite{Breuer2002a}}, but
tends to be awkward if a non-perturbative treatment of the interaction with
the environment is needed.

In the present letter I would like to motivate and exemplify a general method
of obtaining master equations which do incorporate the microscopic
interactions in a non-perturbative fashion. It differs from the standard
approaches in that it takes the Markov assumption not as an approximation in
the course of the calculation, but as a premise, implemented before tracing
out the environment. It will be applicable whenever the interaction with the
environment can reasonably be described in terms of individual interaction
events or ``collisions'', that is, if one can take the environment as
consisting of independent (quasi)-particles which probe the system each at a
time, in the sense that both the rate and the effect of an individual
collision are separately physically meaningful and can be formulated
microscopically. One may then implement the Markov requirement right from the
outset by disregarding the change of the environmental state after each
collision. This will be justified if the environment is sufficiently large and
stationary, and in particular if many different environmental
(quasi)-particles are involved so that each has much time to carry away and
disperse its correlation with the system.

It is clear that the apparatus of time-dependent scattering theory
{\cite{Taylor1972a}} is predestined for this type of description. Its
microscopically defined S-matrix maps from the incoming to the exact outgoing
asymptotes of the system-environment state without a temporal evolution, and a
partial trace over the scattered environment yields the system state after a
single collision. One would like to write the temporal change of the system as
the rate of collisions multiplied by this change due to an individual
scattering. The great difficulty with this is that in general also the rate
depends on the system state so that a naive implementation would yield a
nonlinear equation. Below, I describe how this is circumvented by applying the
concept of generalized and continuous measurements. The use and strength of
the method is then demonstrated by deriving the master equation for the
internal quantum dynamics of an immobile system affected by a gaseous
environment in terms of the multichannel scattering amplitudes.

\section{Monitoring approach}

My first aim is to argue that the system $\rho$ evolves as $\partial_t \rho =
\left( i \hbar \right)^{- 1} \left[ \mathsf{H}, \rho \right] + \mathcal{L}
\rho$ with
\begin{eqnarray}
  \mathcal{L} \rho & = & \frac{i}{2} \tmop{Tr}_{\tmop{env}} \left( \left[
  \mathsf{T} + \mathsf{T}^{\dag}, \Gamma^{1 / 2} \left[ \rho \otimes
  \rho_{\tmop{env}} \right] \Gamma^{1 / 2} \right] \right) \nonumber\\
  &  & + \tmop{Tr}_{\tmop{env}} \left( \mathsf{T} \Gamma^{1 / 2} \left[ \rho
  \otimes \rho_{\tmop{env}} \right] \Gamma^{1 / 2} \mathsf{T}^{\dag} \right)
  \nonumber\\
  &  & - \frac{1}{2} \tmop{Tr}_{\tmop{env}} \left( \Gamma^{1 / 2}
  \mathsf{T}^{\dag} \mathsf{T} \Gamma^{1 / 2} \left[ \rho \otimes
  \rho_{\tmop{env}} \right] \right) \nonumber\\
  &  & - \frac{1}{2} \tmop{Tr}_{\tmop{env}} \left( \left[ \rho \otimes
  \rho_{\tmop{env}} \right] \Gamma^{1 / 2} \mathsf{T}^{\dag} \mathsf{T}
  \Gamma^{1 / 2} \right) .  \label{eq:me1}
\end{eqnarray}
Here $\mathsf{H}$ is the Hamiltonian of the isolated system and
$\rho_{\tmop{env}}$ the reduced single-particle state of the environment. The
operator $\mathsf{T}$ is the nontrivial part of the two-particle S-matrix
$\mathsf{S} = \mathsf{I} + i \mathsf{T}$ describing the effect of a single
collision between environmental particle and system. The rate of collisions is
described by $\Gamma$, a positive operator in the total Hilbert space, which
determines the probability of a collision to occur in a small time interval
$\Delta t$,
\begin{eqnarray}
  \tmop{Prob} \left( \text{C}_{\Delta t} | \rho \right) & = & \Delta t
  \tmop{tr} \left( \Gamma \left[ \rho \otimes \rho_{\tmop{env}} \right]
  \right) .  \label{eq:probscat}
\end{eqnarray}
Like the S-matrix, the operator $\Gamma$ can in principle be characterized
operationally in independent experiments. Its microscopic formulation will in
general involve a total scattering cross section and the current density
operator of the relative motion (see below).

To motivate the time evolution (\ref{eq:me1}) we picture the environment as
monitoring the system continuously by sending probe particles which scatter
off the system at random times. The state dependent collision rate can now be
incorporated into the dynamical description by assuming that the system is
encased by a hypothetical, minimally invasive detector with time resolution
$\Delta t$. It tells at any instant whether a probe particle has passed by and
is going to scatter off the system.

The important point to note is that the information that a collision will
take place changes our description of the impinging two-particle state.
According to the theory of generalized measurements
{\cite{Kraus1983a,Busch1991a,Jacobs2007a}} the new state is the normalized
image of a norm-decreasing completely positive map $\mathcal{M} (\cdot |
\text{C}_{\Delta t})$ in the total Hilbert space satisfying $\tmop{tr} \left(
\mathcal{M} (\varrho | \text{C}_{\Delta t}) \right) = \Delta t \tmop{tr}
\left( \Gamma \varrho \right)$. For an efficient {\cite{efficientmeas}} and
minimally-invasive detector it has the form
\begin{eqnarray}
  \mathcal{M} (\varrho | \text{C}_{\Delta t}) & = & \Delta t \Gamma^{1 / 2}
  \varrho \Gamma^{1 / 2} .  \label{eq:mestra}
\end{eqnarray}
The significance of this measurement transformation is to imprint our improved
knowledge about the incoming two-particle wave packet, and it may be viewed as
enhancing those parts which head towards a collision. In principle, an
efficient measurement (which introduces no classical noise by mapping pure
states to pure states) may be given by a more general operator, $\mathcal{M}
(\varrho | \text{C}_{\Delta t}) = \mathsf{M}_{\Delta t} \varrho
\mathsf{M}_{\Delta t}^{\dag}$ as long as it satisfies $\mathsf{M}_{\Delta
t}^{\dag} \mathsf{M}_{\Delta t} = \Delta t \Gamma$. The above
`minimally-invasive' choice of $\mathsf{M}_{\Delta t}$ is reasonable because a
possible unitary part $\mathsf{U}_{\Delta t}$ in its general polar
decomposition $\mathsf{M}_{\Delta t} = \mathsf{U}_{\Delta t} \Gamma^{1 / 2}
\sqrt{\Delta t}$ would describe a reversible ``back action'' which has no
physical justification in our case of a thought measurement invoked only to
account for the state dependence of collision probabilities.

Also the absence of a detection event during $\Delta t$ changes the state. The
corresponding complementary map $\mathcal{M} (\cdot |
\overline{\text{C}}_{\Delta t})$ satisfies $\tmop{tr} \left( \mathcal{M}
(\varrho | \overline{\text{C}}_{\Delta t}) \right) = 1 - \Delta t \tmop{tr}
\left( \Gamma \varrho \right)$ and the Kraus representation with
time-invariant operators reads $\mathcal{M} (\varrho |
\overline{\text{C}}_{\Delta t}) = \varrho - \Gamma^{1 / 2} \varrho \Gamma^{1 /
2} \Delta t$.

We can now form the unconditioned system-probe state after time $\Delta t$ by
allowing for the fact that the detection outcomes are not really available.
Thus, the infinitesimally evolved state is given by the mixture of the
colliding state transformed by the S-matrix and the untransformed
non-colliding one, weighted with the respective probabilities,
\begin{eqnarray}
  \varrho' \left( \Delta t \right) & = & \tmop{Prob} \left( \text{C}_{\Delta
  t} | \rho \right)  \mathsf{S} \frac{\mathcal{M} (\varrho | \text{C}_{\Delta
  t})}{\tmop{tr} \left( \mathcal{M} (\varrho | \text{C}_{\Delta t}) \right)}
  \mathsf{S}^{\dag} \nonumber\\
  &  & + \tmop{Prob} \left( \overline{\text{C}}_{\Delta t} | \rho \right) 
  \frac{\mathcal{M} (\varrho | \overline{\text{C}}_{\Delta t})}{\tmop{tr}
  \left( \mathcal{M} (\varrho | \overline{\text{C}}_{\Delta t}) \right)}
  \nonumber\\
  & = & \mathsf{S} \Gamma^{1 / 2} \varrho \Gamma^{1 / 2} \mathsf{S}^{\dag}
  \Delta t + \varrho - \Gamma^{1 / 2} \varrho \Gamma^{1 / 2} \Delta t.
  \nonumber
\end{eqnarray}
Using the unitarity of $\mathsf{S}$, which implies $i \left( \mathsf{T} -
\mathsf{T}^{\dag} \right) = - \mathsf{T}^{\dag} \mathsf{T}$, the differential
quotient can be written as
\begin{align}
  \frac{\varrho_{}' \left( \Delta t \right) - \varrho}{\Delta t}  = &
  \mathsf{T} \mathsf{\Gamma}^{1 / 2} \varrho \mathsf{\Gamma}^{1 / 2}
  \mathsf{T}^{\dag} - \frac{1}{2} \mathsf{T}^{\dag} \mathsf{T}
  \mathsf{\Gamma}^{1 / 2} \varrho \mathsf{\Gamma}^{1 / 2} \nonumber\\
    & - \frac{1}{2} \mathsf{\Gamma}^{1 / 2} \varrho \mathsf{\Gamma}^{1 / 2}
  \mathsf{T}^{\dag} \mathsf{T} + i \left[ \tmop{Re} \left( \mathsf{T} \right),
  \mathsf{\Gamma}^{1 / 2} \varrho \mathsf{\Gamma}^{1 / 2} \right] . \nonumber
\end{align}
One arrives at (\ref{eq:me1}) by tracing out the environment with $\varrho =
\rho \otimes \rho_{\tmop{env}}$, taking the limit of continuous monitoring
$\Delta t \rightarrow 0$, and adding the generator of the free system
evolution. Thus, the collision rate with its state dependence is incorporated
by the operators $\Gamma^{1 / 2}$ and they may be thought of, in a stochastic
unravelling of the master equation
{\cite{Carmichael1993a,Gardiner1992a,Molmer1993a,Wiseman1996a,Brun2002a,Cresser2006a}},
as serving to weight each trajectory with the rate before it scatters. The
operators $\mathsf{T}$ describe the individual microscopic interaction process
{\emph{without approximation}}. Note also that (\ref{eq:me1}) generates a
dynamical semigroup by construction since $\mathcal{M} (\cdot |
\text{C}_{\Delta t})$ and $\mathcal{M} (\cdot | \overline{\text{C}}_{\Delta
t})$ are completely positive.

To judge whether the trace in (\ref{eq:me1}) yields a useful master equation
one has to specify system and environment. A first application of this general
equation can already be found in the recent Ref.~{\cite{Hornberger2006b}},
where it is used to describe the motion of a distinguished, freely moving
point-particle in the presence of a gas. The above discussion thus serves to
complete the derivation in {\cite{Hornberger2006b}}, where a quantum version
of the linear Boltzmann equation was obtained which displays all expected
limiting properties. In the following, I will demonstrate the use and
generality of eq.~(\ref{eq:me1}) by posing a complementary question, namely,
how the the {\emph{internal}} dynamics of an immobile system gets affected by
an environment of structureless gas particles.

\section{Application to an immobile system}

If the motional system degrees of freedom are disregarded, a single-particle
S-matrix can be used to describe the (in general inelastic) interaction with
the environmental particles. The resulting master equation should describe
non-perturbatively both the coherent and the incoherent processes induced by
this coupling. An example would be the collisional decay of molecular
eigenstates into chiral configurations, or the phonon-induced decoherence of a
quantum dot. For concreteness, the environment is assumed to be an ideal
Maxwell gas of density $n_{\tmop{gas}}$, atomic mass $m$, and single particle
state ${\rho_{\tmop{env}} = \left( \lambda_{\tmop{th}}^3 / \Omega \right)^{}
\exp \left( - \beta \mathsf{p}^2 / 2 m \right)}$ with $\mathsf{p}$ the
momentum operator, $\lambda_{\tmop{th}} = \hbar \sqrt{2 \pi \beta / m}$ the
thermal wave length, and $\Omega$ the normalization volume.

In the language of scattering theory the free energy eigenstates of the
non-motional degrees of freedom are called channels. In our case, they form a
discrete basis of the system Hilbert space, and $| \alpha \rangle$ will be
used to indicate internal (and possibly rotational), non-degenerate system
eigenstates of energy $E_{\alpha}$. In this channel basis, $\rho_{\alpha
\beta} = \langle \alpha | \rho | \beta \rangle$, the equation of motion
(\ref{eq:me1}) takes on the form of a general master equation of Lindblad
type,
\begin{align}
  \partial_t \rho_{\alpha \beta}  = & \frac{E_{\alpha} + \varepsilon_{\alpha}
  - E_{\beta} - \varepsilon_{\beta}}{i \hbar} \rho_{\alpha \beta} +
  \sum_{\alpha_0 \beta_0} \rho_{\alpha_0 \beta_0} \bignone M_{\alpha
  \beta}^{\alpha_0 \beta_0}  \label{eq:master}\\
  &   - \frac{1}{2}  \sum_{\alpha_0 } \rho_{\alpha_0 \beta_{}} \bignone
  \bignone \sum_{\gamma} M_{\gamma \gamma}^{\alpha_0 \alpha} - \frac{1}{2} 
  \sum_{\beta_0} \rho_{\alpha \beta_0} \bignone \sum_{\gamma} M_{\gamma
  \gamma}^{\beta \beta_0} \nonumber
\end{align}
with energy shifts $\varepsilon_{\alpha}$ discussed below and rate
coefficients
\begin{align}
  M_{\alpha \beta}^{\alpha_0 \beta_0}  = & \langle \alpha |
  \tmop{Tr}_{\tmop{env}} \left( \mathsf{T} \Gamma^{1 / 2} \left[ | \alpha_0
  \rangle \langle \beta_0 | \otimes \rho_{\tmop{env}} \right] \Gamma^{1 / 2}
  \mathsf{T}^{\dag} \right) | \beta \rangle . \nonumber\\
  &    \label{eq:M}
\end{align}
To calculate these complex quantities we need to specify the rate operator
$\Gamma$. In the present case it is naturally given in terms of the current
density operator $\mathsf{j} = n_{\tmop{gas}}  \mathsf{p} / m$ of the
impinging gas particles and the channel-specific total scattering cross
sections $\sigma \left( \tmmathbf{p}, \alpha \right)$,
\begin{eqnarray}
  \Gamma & = & \sum_{\alpha} | \alpha \rangle \langle \alpha | \otimes
  n_{\tmop{gas}}  \frac{\left| \mathsf{p} \right|}{m} \sigma \left(
  \mathsf{p}, \alpha \right) .  \label{eq:Gamma}\\
  &  &  \nonumber
\end{eqnarray}
Defining the channel operator $\mathsf{c} = \sum_{\alpha} \alpha | \alpha
\rangle \langle \alpha |$ one can thus write $\Gamma = | \mathsf{j} | \sigma
\left( \mathsf{p}, \mathsf{c} \right)$.

In principle, $\Gamma$ must also involve a projection to the subspace of
incoming wave packets, attributing zero collision probability to any wave
packet located far off the scattering center and travelling away from it. This
is important because such an outgoing state will not remain invariant under
$\mathsf{S}$. [It may be strongly transformed since the definition of
$\mathsf{S}$ involves a backward evolution.] In practice, the microscopic
definition of $\Gamma$ is easier if one takes care of the projection
separately. This is easily done if $\rho_{\tmop{env}}$ admits a convex
decomposition into incoming and outgoing states. Alternatively, one may
dispense with the projection by modifying the definition of $\mathsf{S}$ so
that outgoing wave packets are kept invariant (see below).

Let us now evaluate the rate coefficients $M_{\alpha \beta}^{\alpha_0
\beta_0}$ by using a decomposition of $\rho_{\tmop{env}}$ that permits to
separate in- and out-wave packets. As shown in {\cite{Hornberger2003b}} the
thermal gas state can be written as a phase space integration over projectors
onto minimum uncertainty gaussian states $| \psi_{\tmmathbf{r}_0
\tmmathbf{p}_0} \rangle = \bar{\lambda}_{\tmop{th}}^{3 / 2} \exp \left( -
\bar{\beta} \left( \mathsf{p} -\tmmathbf{p}_0 \right)^2 / 4 m \right)
|\tmmathbf{r}_0 \rangle$ whose spatial extension $\bar{\lambda}_{\tmop{th}} =
\hbar \sqrt{2 \pi \bar{\beta} / m}$ is determined by an inverse temperature
$\bar{\beta} > \beta$,
\begin{eqnarray}
  \rho_{\tmop{env}} & = & \bignone \int \mathd \tmmathbf{p}_0  \hat{\mu}
  \left( \tmmathbf{p}_0 \right) \int_{\Omega} \frac{\mathd
  \tmmathbf{r}_0}{\Omega} \bignone | \psi_{\tmmathbf{r}_0 \tmmathbf{p}_0}
  \rangle \langle \psi_{\tmmathbf{r}_0 \tmmathbf{p}_0} |.  \label{eq:rhops}
\end{eqnarray}
Here $\hat{\mu} \left( \tmmathbf{p}_0 \right) = (2 \pi m / \hat{\beta})^{- 3 /
2} \exp (- \hat{\beta} \tmmathbf{p}_0^2 / 2 m)$ is the Maxwell-Boltzmann
distribution corresponding to the temperature $\hat{\beta}^{- 1} = \beta^{- 1}
- \bar{\beta}^{- 1}$, so that by setting a $\bar{\beta}$ one splits up the gas
temperature $\beta^{- 1}$ into a part determining the localization of the $|
\psi_{\tmmathbf{r}_0 \tmmathbf{p}_0} \rangle$ and a part characterizing their
motion. We choose $\bar{\beta}$ large and take eventually the limit
$\bar{\beta} \rightarrow \infty$, $\hat{\beta} \rightarrow \beta$ of very
extended wave packets so that $\hat{\mu}$ approaches the original
Maxwell-Boltzmann distribution $\mu$. Inserting (\ref{eq:rhops}) into
(\ref{eq:M}) yields
\begin{eqnarray}
  M_{\alpha \beta}^{\alpha_0 \beta_0} & = & \int \mathd \tmmathbf{p}_0 
  \hat{\mu} \left( \tmmathbf{p}_0 \right) \int_{\Omega} \frac{\mathd
  \tmmathbf{r}_0}{\Omega} \bignone m_{\alpha \beta}^{\alpha_0 \beta_0} \left(
  \tmmathbf{r}_0, \tmmathbf{p}_0 \right) .  \label{eq:M1}
\end{eqnarray}
Here the phase space function
\begin{eqnarray}
  m_{\alpha \beta}^{\alpha_0 \beta_0} \left( \tmmathbf{r}_0, \tmmathbf{p}_0
  \right) & \assign & \int \mathd \tmmathbf{p} \bignone \langle \alpha |
  \langle \tmmathbf{p}| \mathsf{T} \Gamma_{}^{1 / 2} | \alpha_0 \rangle |
  \psi_{\tmmathbf{r}_0 \tmmathbf{p}_0} \rangle \nonumber\\
  &  & \times \langle \beta_0 | \langle \psi_{\tmmathbf{r}_0 \tmmathbf{p}_0}
  | \Gamma_{}^{1 / 2}  \mathsf{T^{\dag}} | \beta \rangle |\tmmathbf{p} \rangle
  \label{eq:m1}
\end{eqnarray}
gives the contribution of different phase space regions to the rate
coefficient $M_{\alpha \beta}^{\alpha_0 \beta_0}$. This permits now to
restrict the calculation to incoming wave packets. Since the $m_{\alpha
\beta}^{\alpha_0 \beta_0}$ are averaged over all available positions in
(\ref{eq:M1}) it is natural to confine this spatial average at fixed
$\tmmathbf{p}_0$ to a cylinder pointing in the direction of $\tmmathbf{p}_0$,
whose longitudinal support $\Lambda_{\tmmathbf{p}_0}$ vanishes at outgoing
positions and whose transverse base area is given by an average cross section
$\Sigma_{\tmmathbf{p}_0}$. In terms of the longitudinal and transverse
positions $\tmmathbf{r}_{\|\tmmathbf{p}_0} \assign \left( \tmmathbf{r} \cdot
\tmmathbf{p}_0 \right) \tmmathbf{p}_0 / p_0^2$ and $\tmmathbf{r}_{\bot
\tmmathbf{p}_0} =\tmmathbf{r}-\tmmathbf{r}_{\|\tmmathbf{p}_0}$ we have
\begin{eqnarray}
  M_{\alpha \beta}^{\alpha_0 \beta_0} & = & \int \mathd \tmmathbf{p}_0 \,
  \hat{\mu} \left( \tmmathbf{p}_0 \right) \int_{\Lambda_{\tmmathbf{p}_0}}
  \frac{\mathd \tmmathbf{r}_{\|\tmmathbf{p}_0}}{\Lambda_{\tmmathbf{p}_0}}
  \bignone \bignone  \int_{\Sigma_{\tmmathbf{p}_0}} \frac{\mathd
  \tmmathbf{r}_{\bot \tmmathbf{p}_0}}{\Sigma_{\tmmathbf{p}_0}} \nonumber\\
  &  & \times m_{\alpha \beta}^{\alpha_0 \beta_0} \left(
  \tmmathbf{r}_{\|\tmmathbf{p}_0} +\tmmathbf{r}_{\bot \tmmathbf{p}_0},
  \tmmathbf{p}_0 \right) .  \label{eq:M2}
\end{eqnarray}
In order to evaluate $m_{\alpha \beta}^{\alpha_0 \beta_0}$, insert momentum
resolutions of unity between the $\mathsf{T}$ and $\Gamma$ operators in
(\ref{eq:m1}) and use the representation {\cite{Taylor1972a}}
\begin{align}
  \langle \alpha_f | \langle \tmmathbf{p}_f | \mathsf{T} | \alpha_i \rangle
  |\tmmathbf{p} \text{$_i$} \rangle  = &  \frac{f_{\alpha_f \alpha_i} \left(
  \tmmathbf{p}_f, \tmmathbf{p}_i \right)}{2 \pi \hbar m}\, \delta \left( E_{p_f
  \alpha_f} - E_{p_i \alpha_i} \right) \nonumber\\
  &    \label{eq:Trep}
\end{align}
in terms of the multi-channel scattering amplitude and total energies $E_{p
\alpha_{}} = p^2 / 2 m + E_{\alpha}$. By transforming the new integration
variables to mid-points and chords one obtains a Gaussian function which
approaches, for large $\bar{\beta}$, a $\delta$-function in the midpoints.
Integrating out the latter one finds that the combination of the
$\delta$-functions from (\ref{eq:Trep}) confine the chord integration to a
plane perpendicular to $\tmmathbf{p}_0$. Integrating out the parallel
component leads to the factor ${\exp \left( - \bar{\beta} m \left( E_{\alpha}
- E_{\alpha_0} - E_{\beta} + E_{\beta_0} \right)^2 / 8 p_0^2 \right)}$ which,
again for large $\bar{\beta}$, can be replaced by
\begin{eqnarray}
  \chi_{\alpha \beta}^{\alpha_0 \beta_0} & \assign & \left\{ \begin{array}{ll}
    1 & \text{if $E_{\alpha} - E_{\alpha_0} = E_{\beta} - E_{\beta_0}$}\\
    0 & \text{otherwise} .
  \end{array} \right. \nonumber
\end{eqnarray}
The resulting expression is independent of $\tmmathbf{r}_{\|\tmmathbf{p}_0}$,
\begin{align}
  m_{\alpha \beta}^{\alpha_0 \beta_0} \left( \tmmathbf{r}_0, \tmmathbf{p}_0
  \right)  = & \chi_{\alpha \beta}^{\alpha_0 \beta_0} 
  \frac{n_{\tmop{gas}}}{m^2} \int \mathd \tmmathbf{p} \bignone \bignone  \int
  \frac{\mathd \tilde{\tmmathbf{p}}_{\bot \tmmathbf{p}_0}}{\left( 2 \pi \hbar
  \right)^2}  \nonumber\\
    & \times \exp \left( - \bar{\beta} \frac{\tilde{\tmmathbf{p}}_{\bot
  \tmmathbf{p}_0}^2}{8 m} - i \frac{\tmmathbf{r}_{0, \bot \tmmathbf{p}_0}
  \cdot \tilde{\tmmathbf{p}}_{\bot \tmmathbf{p}_0}}{\hbar} \right) \nonumber\\
    & \times f_{\alpha \alpha_0} \left( \tmmathbf{p}, \tmmathbf{p}_0^+ 
  \right) f_{\beta \beta_0}^{\ast} \left( \tmmathbf{p}, \tmmathbf{p}_0^-
  \right)  \nonumber\\
    & \times \delta \left( \frac{\tmmathbf{p}^2 - \left( \tmmathbf{p}_0^+
  \right)^2}{2 m} + E_{\alpha} - E_{\alpha_0} \right) \nonumber\\
    & \times \sqrt{\left( 1 + \frac{\tilde{\tmmathbf{p}}_{\bot
  \tmmathbf{p}_0}^2}{4 p_0^2} \right) \sigma \left( \tmmathbf{p}_0^+, \alpha_0
  \right) \sigma \left( \tmmathbf{p}_0^-, \beta_0 \right)} \nonumber
\end{align}
with $\tmmathbf{p}_0^{\pm} \assign \tmmathbf{p}_0 \pm
\tilde{\tmmathbf{p}}_{\bot \tmmathbf{p}_0} / 2$. The
$\tmmathbf{r}_{\|\tmmathbf{p}_0}$-integration in (\ref{eq:M2}) yields an
approximate two-dimensional $\delta$-function in $\tilde{\tmmathbf{p}}_{\bot
\tmmathbf{p}_0}$ so that we obtain
\begin{eqnarray}
  M_{\alpha \beta}^{\alpha_0 \beta_0} & = & \chi_{\alpha \beta}^{\alpha_0
  \beta_0}  \frac{n_{\tmop{gas}}}{m^2} \int \mathd \tmmathbf{p} \bignone
  \bignone \mathd \tmmathbf{p}_0 \mu \left( \tmmathbf{p}_0 \right) f_{\alpha
  \alpha_0} \left( \tmmathbf{p}, \tmmathbf{p}_0  \right) \nonumber\\
  &  & \times f_{\beta \beta_0}^{\ast} \left( \tmmathbf{p}, \tmmathbf{p}_0
  \right) \delta \left( \frac{\tmmathbf{p}^2 -\tmmathbf{p}_0^2}{2 m} +
  E_{\alpha} - E_{\alpha_0} \right), \nonumber\\
  &  &  \label{eq:M3}
\end{eqnarray}
provided we identify the average cross section of (\ref{eq:M2}) with the
geometric mean of the total cross sections of the involved channels, i.e.,
$\Sigma_{\tmmathbf{p}_0} = \sqrt{\sigma \left( \tmmathbf{p}_0 ; \alpha_0
\right) \sigma \left( \tmmathbf{p}_0 ; \beta_0 \right)} $. Moreover, the final
limit $\bar{\beta} \rightarrow \infty$ replaced $\hat{\mu}$ by $\mu$ in
(\ref{eq:M3}).

With the same method one shows that the first term in (\ref{eq:me1}) merely
modifies the unitary evolution. Its effect is to shift the system energies
from $E_{\alpha}$ to $E_{\alpha} + \varepsilon_{\alpha}$ by a thermal average
of the ``forward scattering amplitudes'',
\begin{equation}
  \varepsilon_{\alpha}  =  - 2 \pi \hbar^2 \frac{n_{\tmop{gas}}}{m} \int
  \mathd \tmmathbf{p}_0 \mu \left( \tmmathbf{p}_0 \right) \tmop{Re} \left[
  f_{\alpha \alpha} \left( \tmmathbf{p}_0, \tmmathbf{p}_0 \right) \right] . 
  \label{eq:epsa}
\end{equation}
It is reassuring that the explicit expressions (\ref{eq:M3}) and
(\ref{eq:epsa}) can be shown to be equivalent to the more abstract master
equation by D\"umcke {\cite{Dumcke1985a}}, obtained in a ``low-density limit''
scaling approach {\cite{Breuer2002a,Alicki1987a,Alicki2003a}} for the special
case of a factorizing interaction potential, $\mathsf{V}_{\tmop{tot}} =
\mathsf{A} \otimes \mathsf{B}_{\tmop{env}}$, and for times large compared to
all system time scales. The present approach thus generalizes this result to
arbitrary interaction potentials (satisfying asymptotic completeness) and to
arbitrary times as long as they are greater than the duration of a single
collision.

It is worth noting that the $M_{\alpha \beta}^{\alpha_0 \beta_0}$ can as well
be obtained in a more direct, while less solid way if the {\emph{diagonal}}
momentum representation of $\rho_{\tmop{env}}$ is used instead of
(\ref{eq:rhops}). A projection to the incoming wave packets is then hard to
implement and, as discussed above, the application of $\mathsf{S}$ to
improper 
momentum states leads to the unwanted transformation also of its ``outgoing
components''. As a consequence, the resulting expression for $M_{\alpha
\beta}^{\alpha_0 \beta_0}$ is ill-defined, involving the square of the
$\delta$-functions in (\ref{eq:Trep}) and the normalization volume $\Omega$.
This can be healed by noting that any consistent modification of $\mathsf{S}$
which keeps outgoing wave packets invariant must conserve the probability
current. This condition provides a simple rule how to form a well-defined
expression {\cite{Hornberger2003b,Hornberger2007RR}}, whose multichannel
version yields the result (\ref{eq:M3}) immediately for any momentum diagonal
$\rho_{\tmop{env}}$.

The expression for the rate coefficients can be rewritten, for isotropic
$\mu$, in terms of an average over the velocity distribution $\nu \left( v
\right) = 4 \pi m^3 v^2 \mu \left( mv \right)$ and angular integrations, which
bring about the velocity $v_{\tmop{out}} = \sqrt{v^2 - 2 \left( E_{\alpha} -
E_{\alpha_0} \right) / m}$ of the gas particle after a possibly inelastic
collision. For rotationally invariant scattering amplitudes, $f_{\alpha
\alpha_0} \left( \cos \left( \tmmathbf{p}, \tmmathbf{p}_0 \right) ; E = p_0^2
/ 2 m \right)$, we have
\begin{align}
  M_{\alpha \beta}^{\alpha_0 \beta_0}  = & \chi_{\alpha \beta}^{\alpha_0
  \beta_0} \int_0^{\infty} \mathd v\, \nu \left( v \right) n_{\tmop{gas}}
  v_{\tmop{out}} 2 \pi \int_{- 1}^1 \mathd \left( \cos \theta \right)
  \nonumber\\
    & \times f_{\alpha \alpha_0} \left( \cos \theta ; \frac{m}{2} v^2
  \right) f_{\beta \beta_0}^{\ast} \left( \cos \theta ; \frac{m}{2} v^2
  \right) .  \label{eq:M4}
\end{align}
This shows that limiting cases of (\ref{eq:master}) display the expected
dynamics. For the populations $\rho_{\alpha \alpha}$ it reduces to a rate
equation where the total cross sections $\sigma_{\alpha \alpha_0} \left(
\frac{m}{2} v^2 \right) $ for scattering from channel $\alpha_0$ to $\alpha$
determine the transition rates, $M_{\alpha \alpha}^{\alpha_0 \alpha_0} = \int
\mathd v \nu \left( v \right) n_{\tmop{gas}} v_{\tmop{out}} \sigma_{\alpha
\alpha_0}$. In the case of purely elastic scattering, on the other hand, i.e.,
$M_{\alpha \beta}^{\alpha_0 \beta_0} = M_{\alpha \beta}^{\alpha \beta}
\delta_{\alpha \alpha_0} \delta_{\beta \beta_0}$, the coherences decay
exponentially, $\partial_t \left| \rho_{\alpha \beta} \right| = -
\gamma_{\alpha \beta}^{\tmop{elastic}} \left| \rho_{\alpha \beta} \right|$,
with a rate determined by a {\emph{difference}} of scattering amplitudes,
\begin{align}
  \gamma_{\alpha \beta}^{\tmop{elastic}}  = & \pi \int \mathd v\, \nu \left( v
  \right) n_{\tmop{gas}} v_{\tmop{out}} \int_{- 1}^1 \mathd \left( \cos \theta
  \right) \\
    & \times \left| f_{\alpha \alpha} \left( \cos \theta ; \frac{m}{2} v^2
  \right) - f_{\beta \beta} \left( \cos \theta ; \frac{m}{2} v^2 \right)
  \right|^2 . \nonumber
\end{align}
It shows clearly that the more coherence is lost, in this case, the better the
scattering environment can distinguish between system states $| \alpha
\rangle$ and $| \beta \rangle$.

\section{Conclusions}

In conclusion, a general method of incorporating formal scattering theory into
the dynamic description of open quantum systems was presented. Based on the
theory of generalized measurements, it yields completely positive master
equations which account for the environmental interaction in a
non-perturbative fashion. When applied to an immobile system in the presence
of a gas, it provides a detailed and realistic account of the interplay
between coherent system dynamics and the (possibly much faster) incoherent
effects of the environment.
\acknowledgements 
I thank Bassano Vacchini for helpful discussions. This work was supported by
the DFG Emmy Noether program.

\end{document}